\documentclass[journal]{IEEEtran}
\usepackage{cite}
\usepackage{amsmath,amssymb,amsfonts}
\usepackage{mathtools}
\usepackage{array}
\usepackage{graphicx}
\usepackage{subcaption}
\usepackage{xcolor}
\usepackage[utf8]{inputenc}
\usepackage[T1]{fontenc}
\usepackage[
    colorlinks=true,
    citecolor=blue,
    linkcolor=blue,
    urlcolor=blue
]{hyperref}
\usepackage{orcidlink}

\newcommand{\RLIM}{\mathrm{RLIM}}

\newcommand{\argmax}{\operatorname*{arg\,max}}
\newcommand{\Mult}{\operatorname{Multinomial}}
\newcommand{\Normal}{\mathcal{N}}

\title{Masked Neural Detection for Run-Length-Limited Channel Coding in Molecular Communication}

\author{Melih~\c{S}ahin \orcidlink{0009-0000-6813-3297},~\IEEEmembership{Graduate Student Member,~IEEE,}
Ozgur~B.~Akan \orcidlink{0000-0003-2523-3858},~\IEEEmembership{Fellow,~IEEE}
\thanks{Melih \c{S}ahin and Ozgur B. Akan are with the Centre for neXt Communications (CXC), Department of Engineering, University of Cambridge, CB3 0FA  Cambridge, U.K. (e-mails: \{ms3195, oba21\}@cam.ac.uk).}
\thanks{Ozgur B. Akan is also with the Centre for neXt Communications (CXC), Department of Electrical and Electronics Engineering, Ko\c{c} University, 34450 Istanbul, T\"urkiye.}
\thanks{This work was supported in part by the AXA Research Fund (AXA Chair for Internet of Everything at Ko\c{c} University).}
}

\begin{document}
\maketitle

\begin{abstract}
Molecular communication (MC) suffers from severe diffusion memory because molecules released for one symbol may arrive during later symbol intervals. Neural sequence detectors, especially sliding bidirectional recurrent neural networks (SBRNNs), substantially outperform threshold detection in such channels. This raises a central question for MC channel coding: does a code whose superiority was established under threshold detection retain it when both coded and uncoded transmission are evaluated with neural detection? This letter answers this question for run-length-limited ISI-mitigation (RLIM) codes by proposing a decoder-aware training mask that removes the positions the RLIM decoder has a high probability of deterministically overwriting, steering compact-SBRNN capacity toward the information-bearing positions. The masked RLIM$_2$-SBRNN beats the best uncoded receiver (threshold or SBRNN) at 40 of 57 operating points; gains peak at 43$\times$ under favorable channels, while losses, confined to the most adverse, never exceed 2.7$\times$. Masking improves the unmasked RLIM$_2$-SBRNN in 56 of 57 matched comparisons. Finally, with storage counted equally in SBRNN weights and MLSE table entries, the masked RLIM$_2$-SBRNN is the more accurate receiver up to a few thousand stored values despite using no channel knowledge; channel-state-aware uncoded MLSE moves ahead only beyond tens of thousands.
\end{abstract}


\section{Introduction}

Molecular communication (MC) conveys data by emitting molecules that diffuse from a transmitter to a receiver, and is studied for nanoscale and in-body channels where conventional electromagnetic methods are hard to deploy \cite{Akan2017,Kuscu2019}, with applications from in-body sensing and drug delivery to Internet of Bio-Nano Things interfaces. In diffusion-based MC, the molecules emitted for one symbol do not all arrive within that symbol time interval: some are captured at once, others during later intervals, and a fraction never. The late arrivals interfere with later symbols, so the count in any interval depends on several past symbols, making reliable reception a sequence-detection problem.

The simplest MC receivers threshold each received count, ignoring the channel memory that diffusion creates \cite{Kuscu2019}. Detectors that exploit this memory optimally need channel state information (CSI) (i.e., diffusion coefficient, geometry, and noise level) which is hard to obtain on a small MC node \cite{Kilinc2013Receiver}; a data-driven detector instead learns the count-to-bit mapping from data. The sliding bidirectional recurrent neural network (SBRNN) suits this task \cite{FarsadGoldsmith2018}: a bidirectional LSTM runs over a short window of received counts, producing a per-bit probability mass function (PMF), and as the window slides each bit is covered by many windows whose PMFs are averaged. In our sweeps, the uncoded SBRNN and the proposed masked coded SBRNN outperform their optimized threshold counterparts in bit error rate (BER) at every operating point tested.

Interference can also be reduced at the transmitter by constraining which bit patterns are sent \cite{Kislal2020,SahinAkan2024RLIM}. Run-length-limited ISI-mitigation (RLIM) coding does this \cite{SahinAkan2024RLIM}: from the run-length-limited words of a given length it keeps the $2^k$ of lowest Hamming weight, so the transmitted $1$-bits are few and well separated and, under the standard MC power normalization, each $1$-bit is sent with more molecules. RLIM outperforms classical run-length-limited (RLL) and other prominent MC codes over a wide range of conditions \cite{SahinAkan2024RLIM}, and a later enumerative realization replaced its exponentially growing full-codebook storage with polynomial-size counting tables \cite{SahinAkan2026LowStorage}.

The central contribution of this letter is to show that a constrained code can reshape a neural receiver's learning objective, and to realize it by training the SBRNN detector \cite{FarsadGoldsmith2018} under our proposed mask, dictated by the RLIM decoder \cite{SahinAkan2024RLIM}. In RLIM$_i$, every detected triggering $1$-bit makes the decoder overwrite the following $i$ positions with zeros, yet standard cross-entropy still penalizes the network at these forced-zero positions, which are unusually difficult because residual molecules from the preceding emission make their counts resemble $1$-bits. An unmasked compact network is thus trained to solve a hard task whose output the decoder discards; our mask removes exactly those gradients and steers the limited capacity toward triggering $1$-bits and unconstrained zeros. Section~\ref{sec:results} quantifies the benefit of this reallocation with position-level
error diagnostics.

Neural detectors and decoders for constrained sequences exist outside MC, but with different training targets. In~\cite{Zheng2020PRNN}, an RNN recovers the constrained channel-input symbols of a partial-response channel, which a sliding-block decoder then maps to the user data; the detector is trained on all of those symbols, excluding none from its loss. In~\cite{Cao2019CSDecoding}, for fixed-length constrained codes, the network replaces the decoder and is trained to output the coded bits directly, so there are no code-determined positions to remove. Our mask applies whenever a decoder deterministically overwrites specific positions once a triggering $1$-bit is detected; the present hard mask does not apply without such positions. To our knowledge, this is the first neural detector for constrained-coded diffusion-based MC; unlike prior neural constrained-sequence methods, its training loss drops the data-dependent positions a decoder overwrites, rather than fixed positions a code freezes, as with polar frozen bits~\cite{Cammerer2017Partition}.

The remainder of the paper is organized as follows. Section~\ref{sec:model} gives the MC channel model and the RLIM code. Section~\ref{sec:sbrnn} presents the SBRNN detector and the proposed mask. Section~\ref{sec:results} reports the simulation framework, repeated-run uncertainty, triggering-$1$ error analysis, and the storage-matched MLSE comparison. Section~\ref{sec:conclusion} concludes the paper.

\section{MC Channel Model}
\label{sec:model}

\begin{figure}[!t]
\centering
\includegraphics[width=\columnwidth]{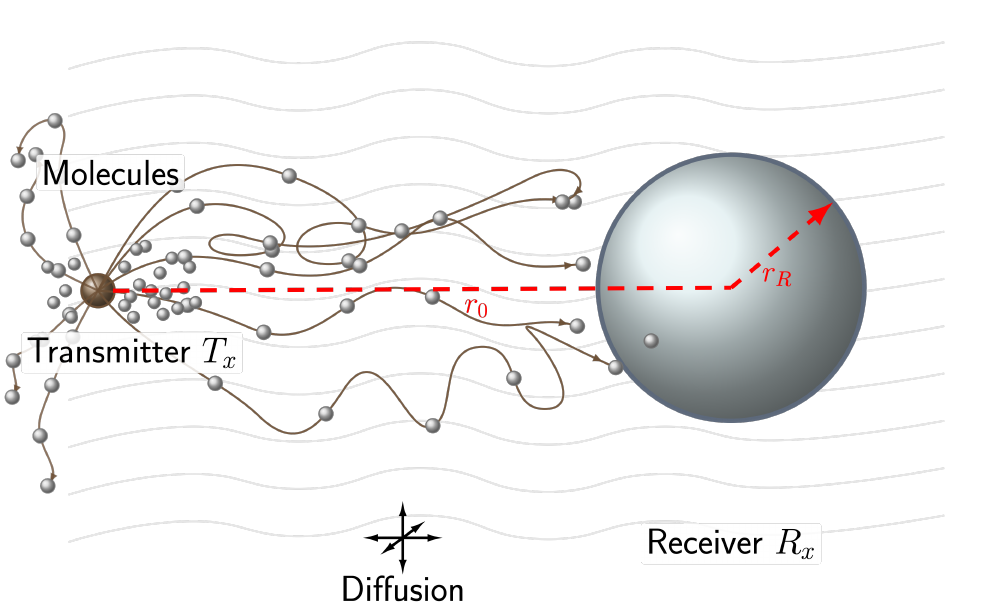}
\caption{Diffusion-based MC channel model, following \cite{SahinAkan2026LowStorage}.}
\label{fig:channel}
\end{figure}

We consider a diffusion-based MC system in which a single point transmitter communicates with a single fully absorbing spherical receiver by binary concentration-shift keying \cite{BCSK}, as shown in Fig.~\ref{fig:channel}. The transmitter sits at distance $r_0$ from the centre of the receiver, the receiver has radius $r_R$, and the medium has diffusion coefficient $D$. A $1$-symbol is represented by an ideal instantaneous release (a Dirac impulse in emission time) of a number $M$ of molecules at the start of the symbol interval; a $0$-symbol releases none. The probability that a single molecule has been absorbed by time $t$ is \cite{Yilmaz2014Absorb}
\begin{equation}
F(t)=
\frac{r_R}{r_0}
\operatorname{erfc}\!\left(
\frac{r_0-r_R}{\sqrt{4Dt}}
\right) .
\label{eq:F}
\end{equation}
With a signaling interval of length $t_s$, a molecule emitted at the start of an interval reaches the receiver in the $\ell$-th interval that follows with probability \cite{Kuscu2019}
\begin{equation}
p_\ell=F(\ell t_s)-F((\ell-1)t_s),
\qquad \ell=1,\ldots,I,
\label{eq:taps}
\end{equation}
where $I$ is the retained channel memory, and the fraction not absorbed within these $I$ intervals is $\pi_{\mathrm{tail}}=1-\sum_{\ell=1}^{I}p_\ell$. A single emission of $M$ molecules is therefore split among these $I+1$ outcomes by a multinomial law \cite{multinomial},
\begin{equation}
(X_1,\ldots,X_I,X_{I+1})
\sim
\Mult(M;p_1,\ldots,p_I,\pi_{\mathrm{tail}}).
\label{eq:mult}
\end{equation}
The count measured in interval $t$ then collects the contributions of all earlier emissions that reach it \cite{SahinAkan2026LDP}, together with receiver noise,
\begin{equation}
N_t=
\sum_{\ell=1}^{\min\{t,I\}}
c_{t-\ell+1}X_{\ell}^{(t-\ell+1)}
+
W_t,
\label{eq:received}
\end{equation}
where $c_t\in\{0,1\}$ is the encoded bit, $X_{\ell}^{(s)}$ is the number of molecules from the emission at time $s$ that arrive $\ell$ intervals later, and $W_t\sim\lfloor\Normal(0,\sigma^2)\rceil$ is the rounded counting noise of the receiver. This multinomial model is used in all of our simulations.

We now define the RLIM codes \cite{SahinAkan2024RLIM}. Fix an order $i$ and a block length $n$, and let $\mathcal{C}_i(t)\subseteq\{0,1\}^t$ be the length-$t$ binary words in which every pair of consecutive $1$-bits is separated by at least $i$ zeros, i.e., the $(i,\infty)$-run-length-limited words. Prefixing $i$ zeros to each word of $\mathcal{C}_i(n-i)$ gives the enhanced (i.e., containing the all-zero code) RLIM family of length $n$,
\begin{equation}
\widehat{\RLIM}_i(n)=\bigl\{\,0^{\,i}\!\circ u \;:\; u\in\mathcal{C}_i(n-i)\,\bigr\}.
\label{eq:rlim_family}
\end{equation}
 From this family we keep the $2^k$ words of smallest Hamming weight, breaking ties lexicographically, and map the $k$ information bits onto them \cite{SahinAkan2024RLIM,SahinAkan2026LowStorage}. The shortest admissible block length is $n=\min\{\,n\ge i+1:|\mathcal{C}_i(n-i)|\ge 2^k\,\}$. The all-zero word is retained because the receiver uses a static threshold or a neural detector, not a dynamic threshold.

Schemes are compared at matched power and rate. Let $M_0$, $t_{s,0}$, and $I_0$ be the molecule budget, signaling interval, and retained channel-memory interval count of the uncoded scheme. A scheme with $k$ information bits, block length $n$, and total codebook $1$-bit count $W$ uses
\begin{equation}
\begin{aligned}
t_s&=t_{s,0}\frac{k}{n}, &
M&=\operatorname{round}\!\left(M_0\frac{W_0}{W}\right),\\
I&=\left\lceil I_0\frac{n}{k}\right\rceil, &
W_0&=k\,2^{k-1},
\end{aligned}
\label{eq:normalization}
\end{equation}
so that one coded block occupies the time of $k$ uncoded bits, spends about the same number of molecules per codebook, and retains at least the same physical diffusion-memory duration $I_0t_{s,0}$ \cite{SahinAkan2024RLIM,SahinAkan2026LDP}. Table~\ref{tab:rlim_normalization} lists the normalization factors for $k=16$. For the comparisons in this letter, we use RLIM$_2$ as a representative intermediate-order code, which gives greater $1$-bit separation than RLIM$_1$ while retaining a higher coding rate than RLIM$_3$ and RLIM$_4$ \cite{SahinAkan2024RLIM}.

\begin{table}[!t]
\centering
\caption{Enhanced RLIM parameters and normalization factors for $k=16$.}
\label{tab:rlim_normalization}
\renewcommand{\arraystretch}{0.96}
\setlength{\tabcolsep}{3.5pt}
\footnotesize
\begin{tabular}{lccccc}
\hline
Scheme & $n$ & $k/n$ & $W$ & $M/M_0$ & $t_s/t_{s,0}$ \\
\hline
Uncoded & 16 & 1.000 & 524288 & 1.000 & 1.000 \\
RLIM$_2$ & 31 & 0.516 & 353221 & 1.484 & 0.516 \\
\hline
\end{tabular}
\end{table}

Detection and decoding are two stages. The detector first maps the length-$n$ vector of received counts $N\in\mathbb{Z}^n$ to a binary word $b\in\{0,1\}^n$ (by thresholding, or by the neural detector of Section~\ref{sec:sbrnn}). The decoder then turns $b$ into a codeword in two steps. A left-to-right pass restores the run-length structure: it sets the first $i$ positions to $0$, and thereafter sets $c_t=1$ only if $b_t=1$ and no $1$-bit has been kept within the previous $i$ positions, and $c_t=0$ otherwise, so each kept $1$-bit opens a window of $i$ forced $0$-bits and any $1$-bit inside such a window is deleted. If the corrected word $c$ is not among the $2^k$ selected codewords, projection decoding clears its rightmost $1$-bits one at a time until $c$ re-enters the codebook \cite{SahinAkan2024RLIM}; the $k$ information bits are then read from the rank of $c$ in the weight-ordered family. For example, with $i=2$ a detected block
$b=0\,1\,0\,1\,1\,0\,1\,0$ corrects to
$c=0\,0\,0\,1\,0\,0\,1\,0$
(positions 2 and 5 deleted; positions 4 and 7 kept; positions 5, 6, 8 forced to 0),
which projection then maps to a selected codeword if it is not already one.

\section{SBRNN Detection and Proposed Masking}
\label{sec:sbrnn}

We detect the encoded bits $(c_t)_{t=1}^{T}$ from the count sequence with the SBRNN of~\cite{FarsadGoldsmith2018}. Let $T$ be the number of received counts, and let $L$ be the window length. A window starting at position $s$ holds the counts $(N_s,\ldots,N_{s+L-1})$, and a bidirectional LSTM maps it to a posterior over the transmitted bit at each offset $u\in\{0,\ldots,L-1\}$,
\begin{equation}
q_{s,u}(a)
=
\Pr_\theta(c_{s+u}=a\mid N_s,\ldots,N_{s+L-1}),
\qquad a\in\{0,1\}.
\label{eq:sbrnn_window_pmf}
\end{equation}
Here $q_{s,u}\colon\{0,1\}\to[0,1]$ is a probability mass function (a softmax output, so $q_{s,u}(0)+q_{s,u}(1)=1$), and $q_{s,u}(a)$ denotes the scalar probability it assigns to the bit value $a$. As the window advances one position at a time, bit $t$ is covered by the windows $\mathcal{S}(t)=\{s:1\le s\le T-L+1,\;s\le t\le s+L-1\}$, whose posteriors are averaged \cite{FarsadGoldsmith2018},
\begin{equation}
\bar q_t(a)
=
\frac{1}{|\mathcal{S}(t)|}
\sum_{s\in\mathcal{S}(t)}
q_{s,t-s}(a),
\qquad
\widehat c_t=
\argmax_{a\in\{0,1\}}\bar q_t(a).
\label{eq:sbrnn_avg}
\end{equation}
The hard bits $(\widehat c_t)$ are handed to the same RLIM decoder of Section~\ref{sec:model}; the SBRNN simply replaces thresholding as the count-to-bit stage.

A general-purpose SBRNN can carry hundreds of thousands of parameters, undesirable for the small devices expected in micro- and nanoscale MC \cite{Kuscu2019}. We use a one-layer bidirectional LSTM with hidden size $H=16$ (2498 parameters) as the default. The choice is empirical: in the capacity sweep of Fig.~\ref{fig:results}(g), the masked RLIM$_2$-SBRNN BER falls from $5.33\times10^{-5}$ at 242 parameters to $3.84\times10^{-5}$ at 2498 and is then approximately flat within run-to-run spread up to 363202 parameters, so 2498 gives the most compact model that reaches the performance floor. The window $L=40$ is read off the context sweep in Fig.~\ref{fig:results}(f). For input size one, hidden size $H$, $R$ layers, and a two-class output, the trainable-parameter count is
\begin{equation}
P(H,R)=
\sum_{r=1}^{R}
2\bigl(4H d_r+4H^2+8H\bigr)
+
2(2H)+2,
\label{eq:param_count}
\end{equation}
with $d_1=1$ and $d_r=2H$ for $r\ge 2$. Table~\ref{tab:sbrnn_architectures} lists the architectures.

\begin{table}[!t]
\centering
\caption{SBRNN architectures used in the storage sweep.}
\label{tab:sbrnn_architectures}
\renewcommand{\arraystretch}{1.04}
\setlength{\tabcolsep}{5.0pt}
\footnotesize
\begin{tabular}{lccc}
\hline
Role & Hidden size & Layers & Parameters \\
\hline
Tiny & 4 & 1 & 242 \\
Small & 8 & 1 & 738 \\
Default compact & 16 & 1 & 2498 \\
Moderate & 32 & 1 & 9090 \\
High-capacity reference & 80 & 3 & 363202 \\
\hline
\end{tabular}
\end{table}

The default's 2498 weights occupy $9.76$~KiB in FP32 as trained here, and would fall to about $2.44$~KiB under 8-bit quantization. Realizing the detector nonetheless presumes a digital processing and memory unit alongside the receiver: the artificial MC receivers surveyed in~\cite{Kuscu2019} are biochemical sensing front-ends, such as FET-based biosensors, without on-node digital computation, so integrating a learned detector, particularly within a resource-constrained biological node, remains future work.

The proposed method changes only training. An order-$i$ RLIM code forces $i$ $0$-bits after every $1$-bit, and we exclude exactly these positions from the loss. Let $m_t$ be the mask
\begin{equation}
m_t=
\begin{cases}
0, & c_t \text{ is one of the } i \text{ forced } 0\text{-bits after a } 1\text{-bit},\\
1, & \text{otherwise}.
\end{cases}
\label{eq:mask_definition}
\end{equation}
Let $\mathcal{W}$ be the set of window-start positions in the labelled training sequence. For a window $s\in\mathcal{W}$ and offset $u$ (absolute position $t=s+u$), the PMF in Eq.~\eqref{eq:sbrnn_window_pmf} assigns the scalar probability $q_{s,u}(c_{s+u})\in[0,1]$ to the true bit $c_{s+u}\in\{0,1\}$, giving the per-position cross-entropy $-\ln (q_{s,u}(c_{s+u}))\ge 0$. With the mask $m_t$ in Eq.~\eqref{eq:mask_definition}, the detector is trained by minimizing the mask-weighted mean cross-entropy
\begin{equation}
\mathcal{L}(\theta)
=
\frac{
\sum_{s\in\mathcal{W}}\;\sum_{u=0}^{L-1}
m_{s+u}\,\bigl(-\ln (q_{s,u}(c_{s+u}))\bigr)
}{
\sum_{s\in\mathcal{W}}\;\sum_{u=0}^{L-1}m_{s+u}
}.
\label{eq:masked_loss}
\end{equation}

The mask is used only in training. At test time the network scores every position, and the RLIM decoder then restores the run-length constraint, so the decision rule in Eq.~\eqref{eq:sbrnn_avg} is unchanged. Residual molecules make the forced-zero positions resemble $1$-bits. An unmasked network therefore wastes capacity on positions the decoder discards, so masking helps most for compact models, as Fig.~\ref{fig:results}(g) shows. The mask assumes triggering $1$-bits are usually detected correctly, which Table~\ref{tab:circularity} confirms. We now examine these points in detail.

\begin{figure*}[!t]
\centering
\newlength{\HOVERLAP}\setlength{\HOVERLAP}{3mm}
\newlength{\PANELW}\setlength{\PANELW}{\dimexpr 0.24\textwidth + 0.75\HOVERLAP \relax}
\newlength{\ROWGAP}\setlength{\ROWGAP}{0.6mm}
\setlength{\tabcolsep}{0pt}
\begin{tabular}{@{}c@{\hspace{-\HOVERLAP}}c@{\hspace{-\HOVERLAP}}c@{\hspace{-\HOVERLAP}}c@{}}
\multicolumn{4}{@{}c@{}}{\includegraphics[width=1\textwidth]{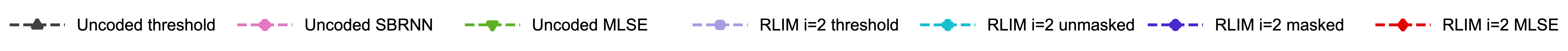}}\\[-0.5mm]
\subcaptionbox{$t_{s,0}=0.2$~s}{\includegraphics[width=\PANELW]{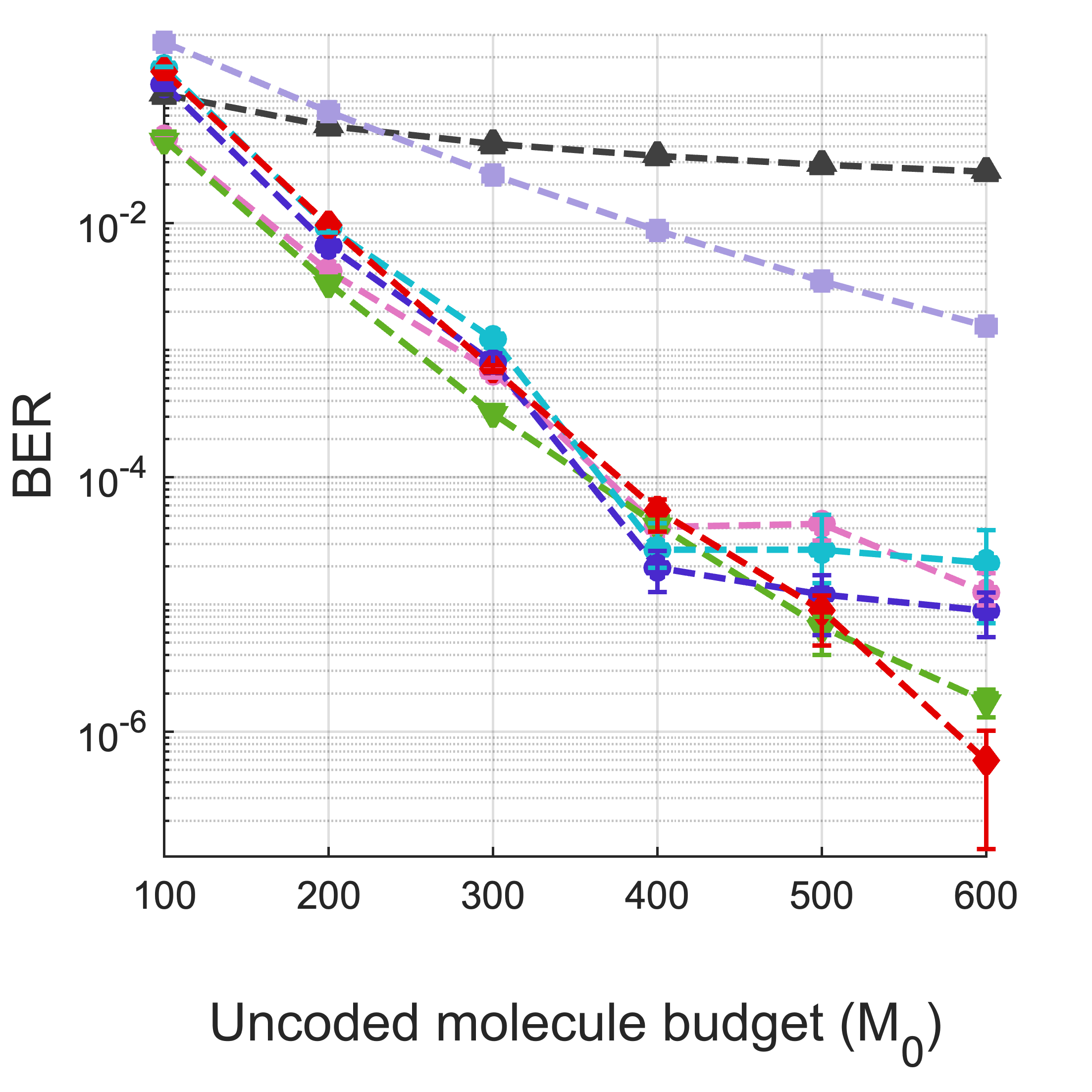}} &
\subcaptionbox{$M_0=100$}{\includegraphics[width=\PANELW]{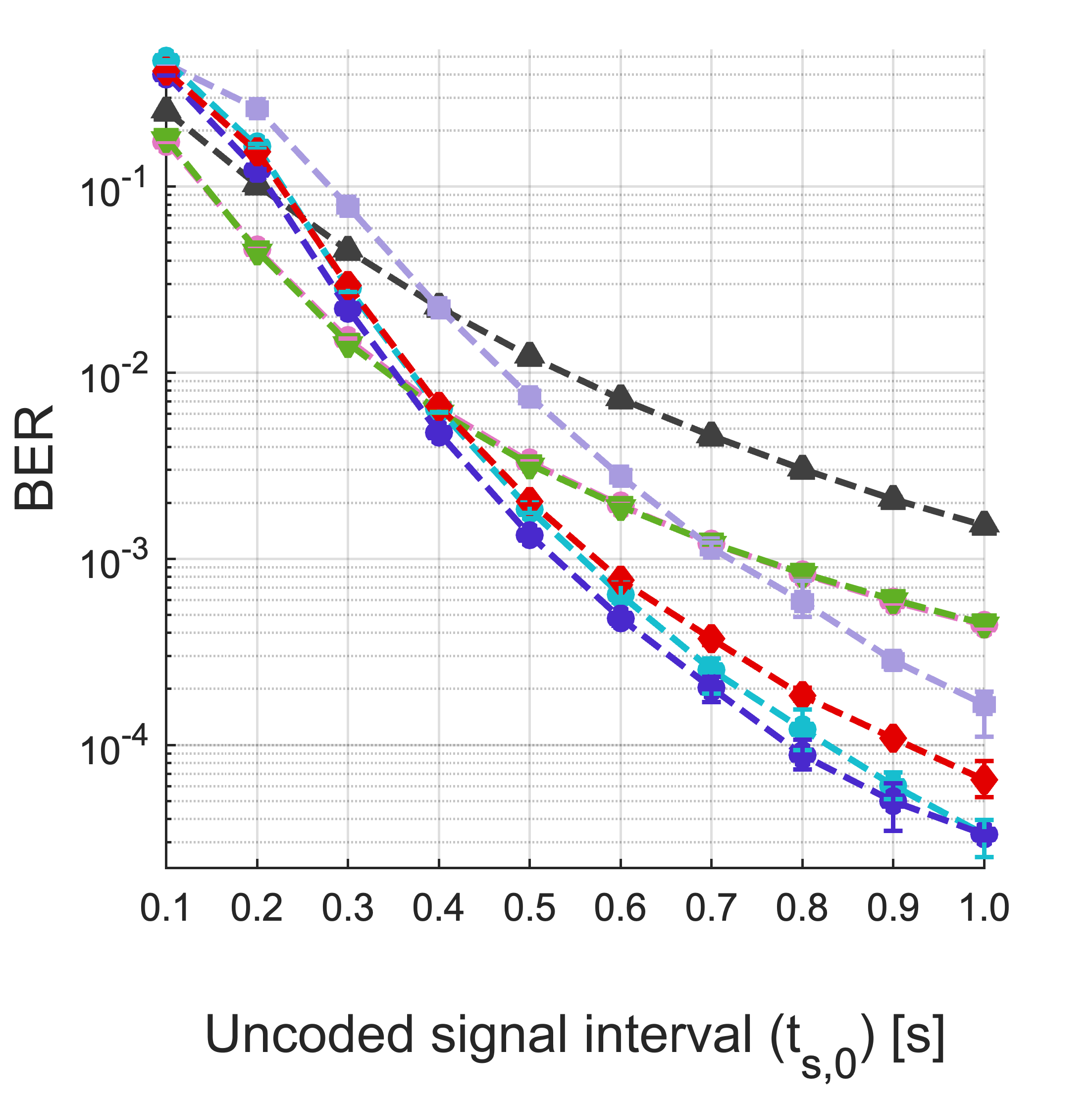}} &
\subcaptionbox{$M_0=400$, $t_{s,0}=0.2$~s}{\includegraphics[width=\PANELW]{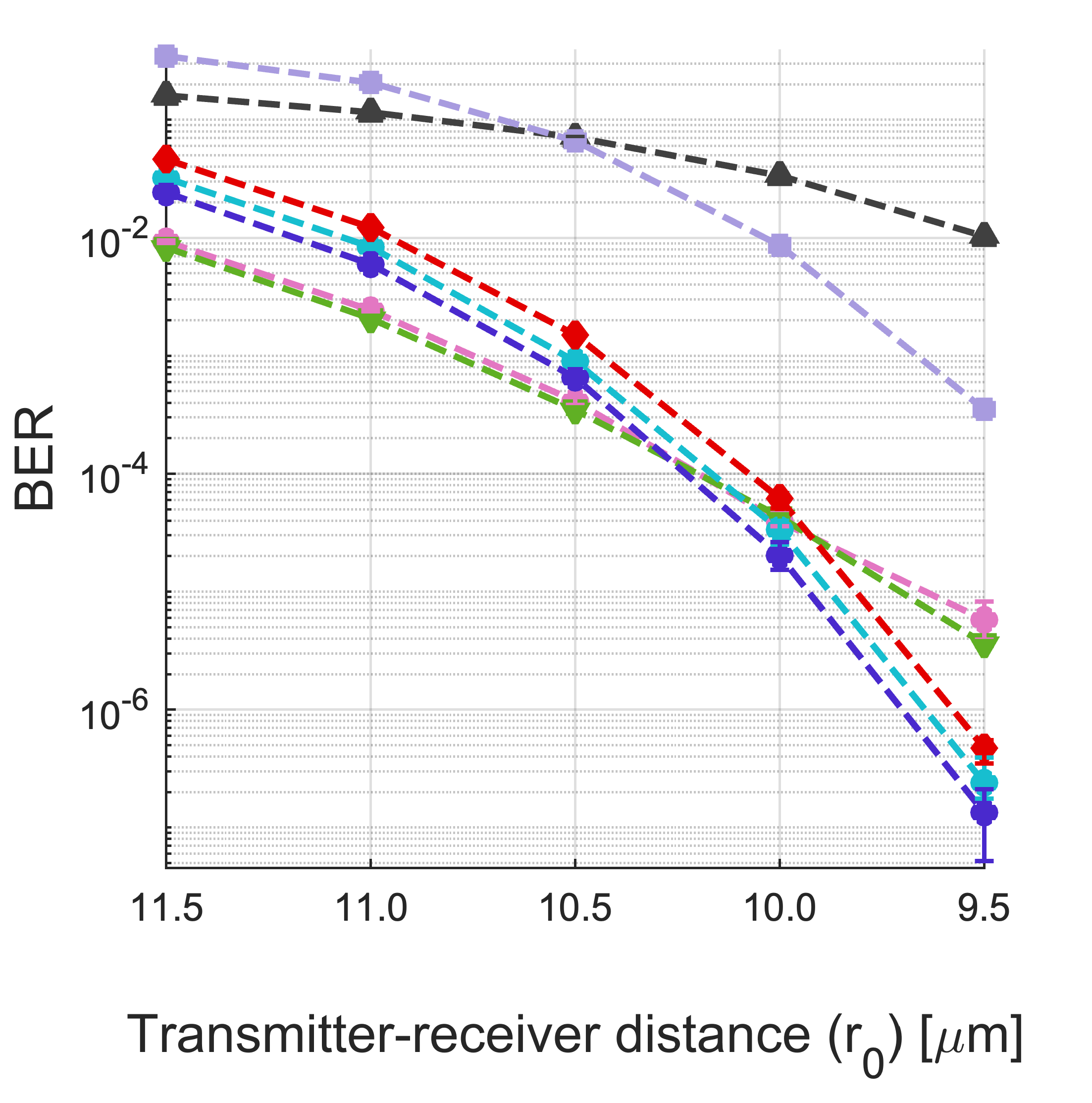}} &
\subcaptionbox{$M_0=400$, $t_{s,0}=0.2$~s}{\includegraphics[width=\PANELW]{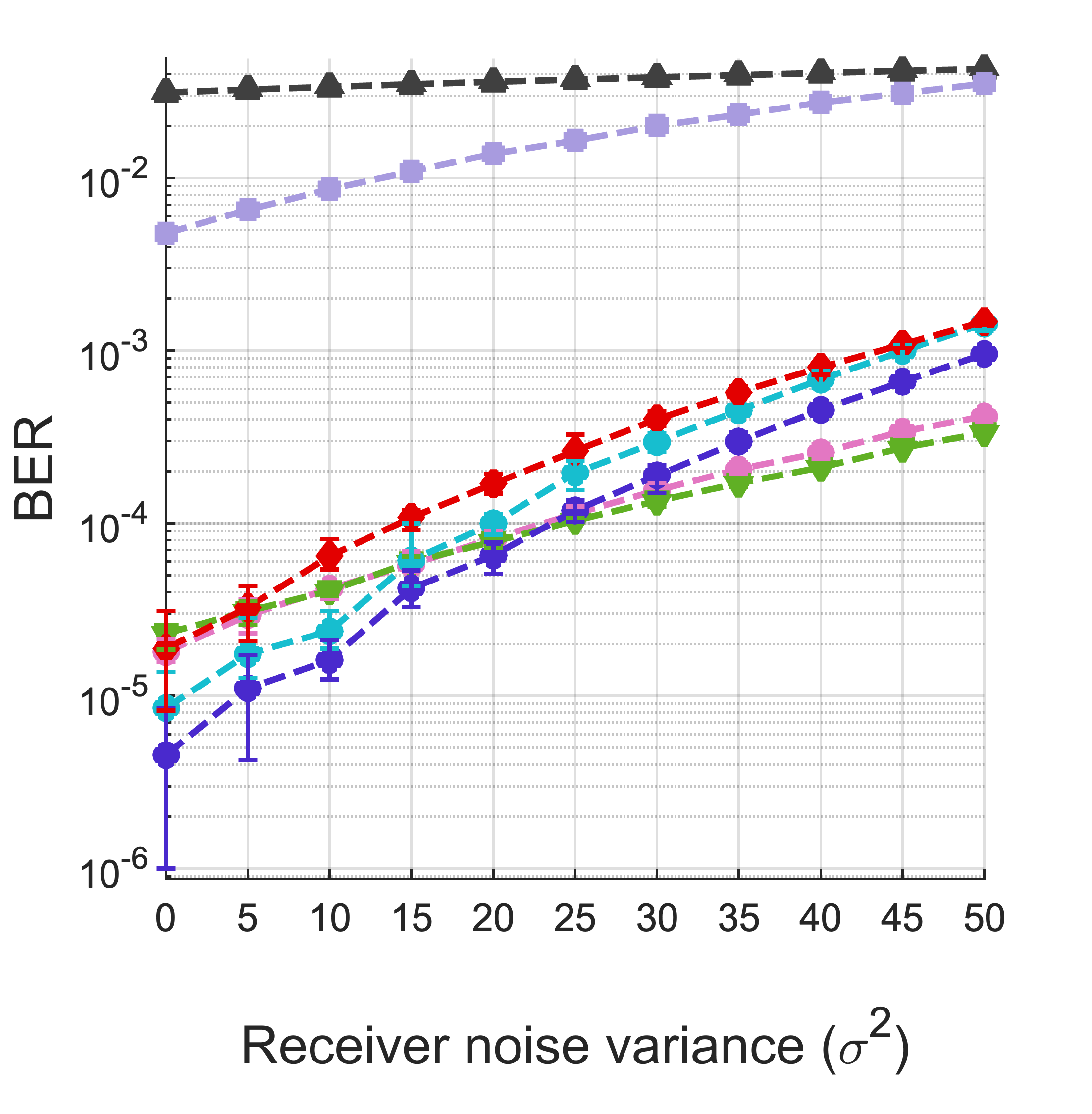}} \\[\ROWGAP]
\subcaptionbox{$t_{s,0}=0.2$~s}{\includegraphics[width=\PANELW]{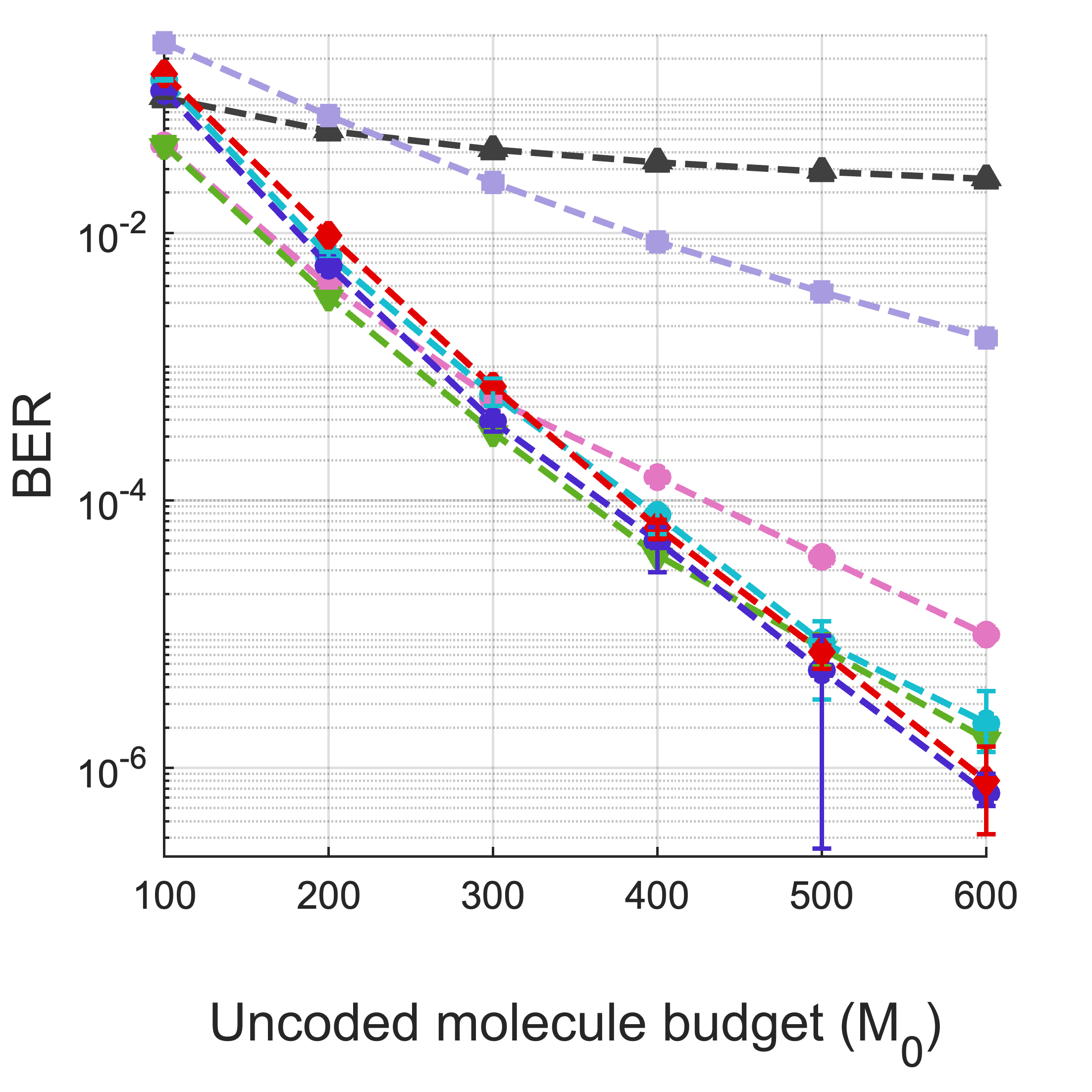}} &
\subcaptionbox{$M_0=400$, $t_{s,0}=0.2$~s}{\includegraphics[width=\PANELW]{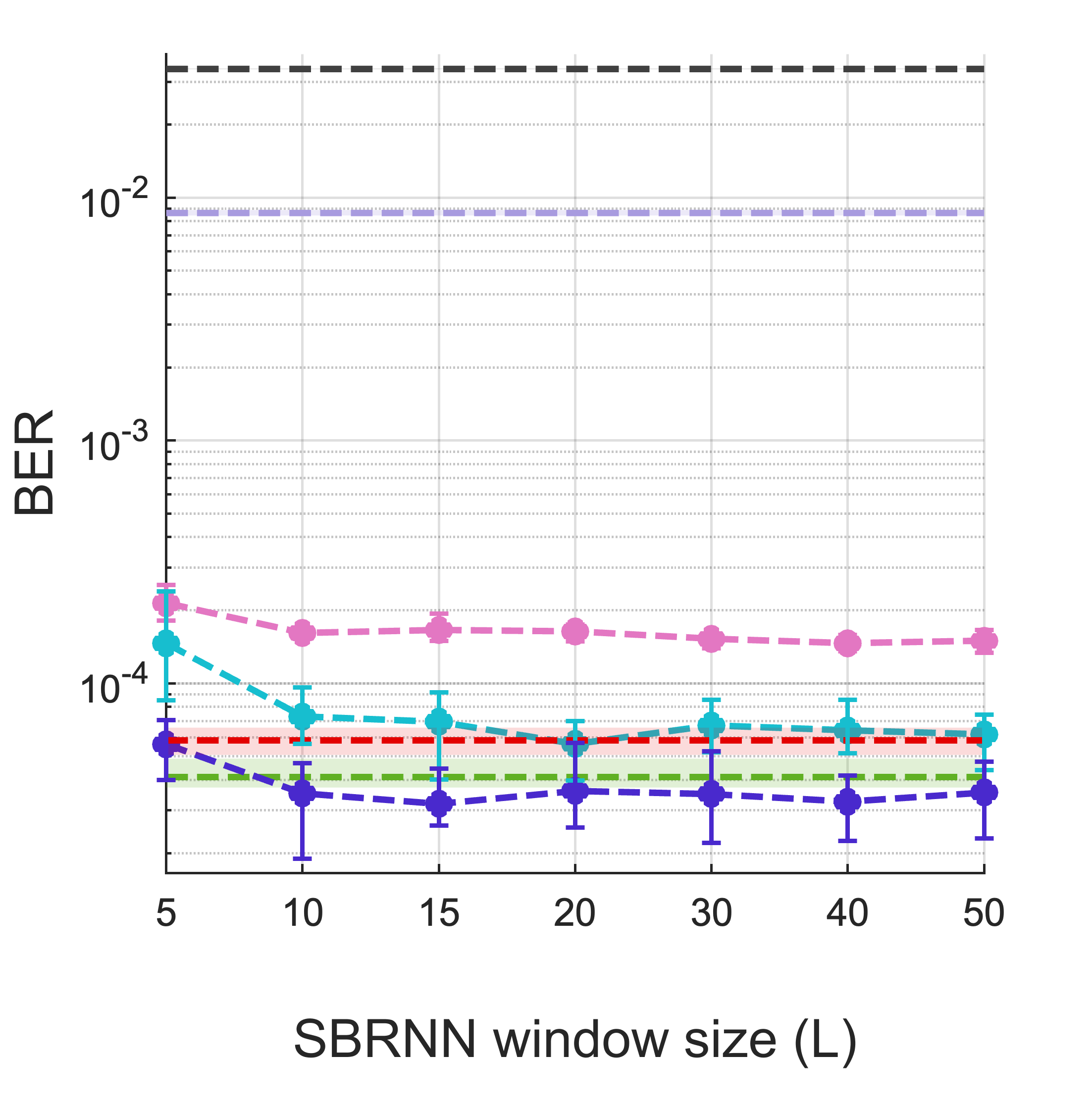}} &
\subcaptionbox{$M_0=400$, $t_{s,0}=0.2$~s}{\includegraphics[width=\PANELW]{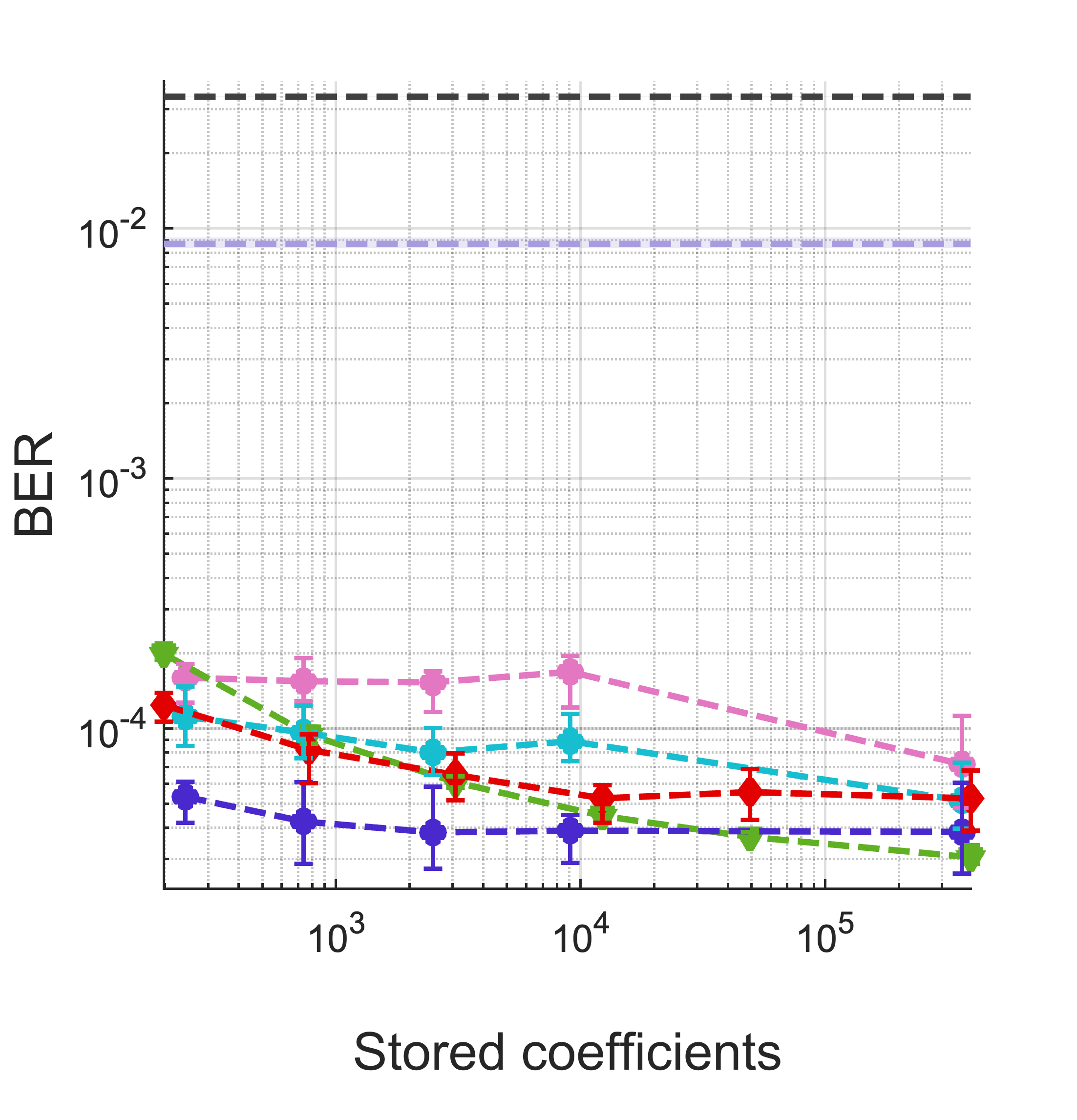}} &
\subcaptionbox{$M_0=400$, $t_{s,0}=0.2$~s}{\includegraphics[width=\PANELW]{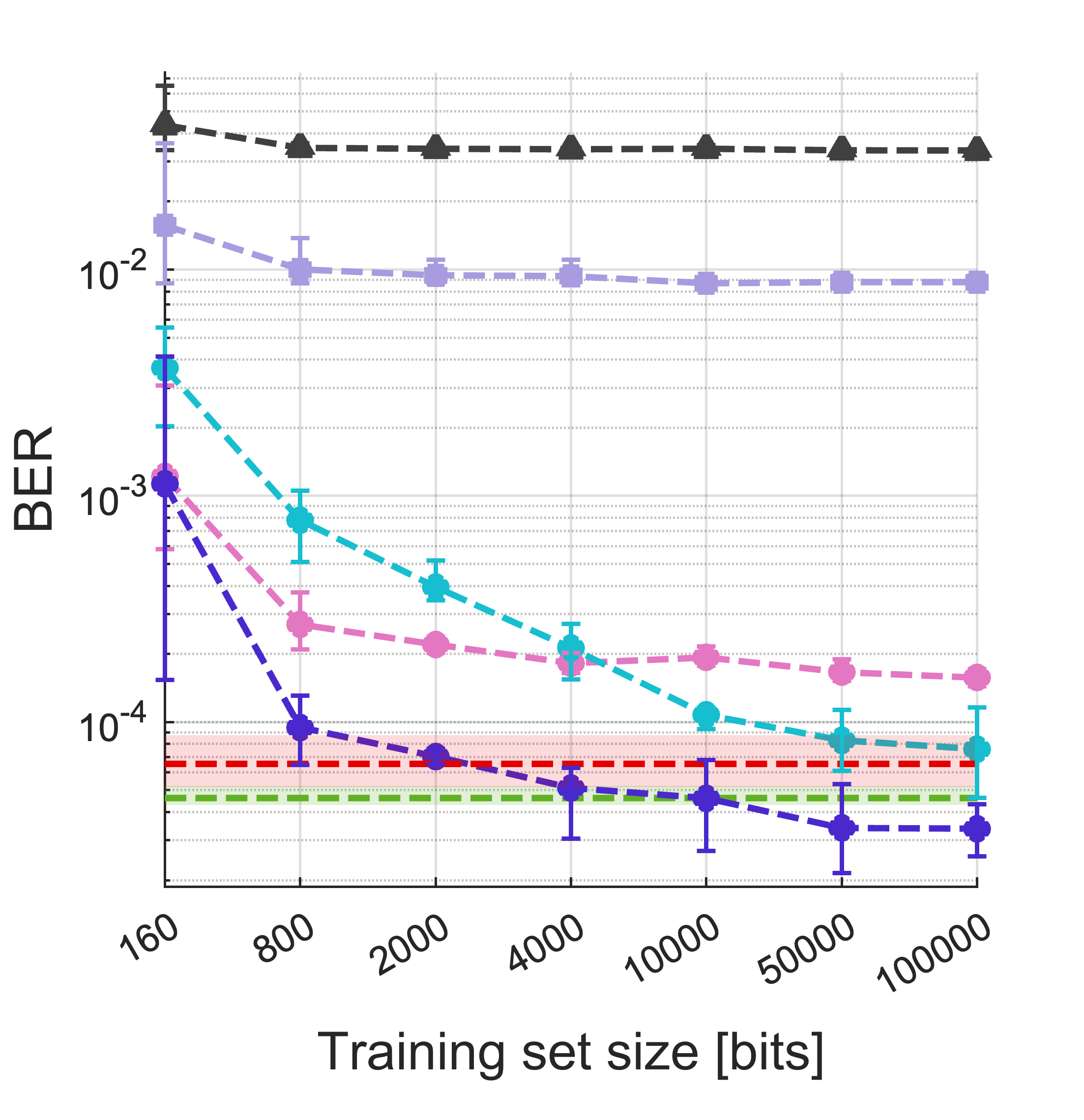}} \\
\end{tabular}
\caption{Mean BER over five independent training/test replicates; error bars span the minimum and maximum BER across the five runs, and each subtitle lists only the channel parameters that differ from the common settings in the text. Panels (a)--(d) reuse one early-stopped pooled cross-condition SBRNN per method/loss; panels (e)--(h) train a separate early-stopped SBRNN at each point.}
\label{fig:results}
\end{figure*}

\section{Simulation Framework and Results}
\label{sec:results}

We evaluate seven receivers: the uncoded threshold, SBRNN, and reduced-state MLSE, together with their RLIM$_2$ counterparts, i.e., the RLIM$_2$ threshold, unmasked SBRNN, masked SBRNN, and reduced-state MLSE. We fix the code dimension $k=16$, the noise variance $\sigma^2=10$, and the retained channel memory of the uncoded scheme at $I_0=100$ intervals, so that $I=\lceil I_0 n/k\rceil=194$ for RLIM$_2$. The physical channel constants follow a standard insulin-based MC model.\footnote{$D=79.4~\mu\mathrm{m}^2/\mathrm{s}$, $r_R=5~\mu\mathrm{m}$, and $r_0=10~\mu\mathrm{m}$, modeling human insulin as the information molecule~\cite{ISI_mitigating_methods_2015}.} The simulator is written in Python, using NumPy for the channel and PyTorch for the SBRNN; with the mask of Eqs.~\eqref{eq:mask_definition}--\eqref{eq:masked_loss} and the MLSE described below, Section~\ref{sec:model} specifies the full pipeline for reproduction. Each plotted point is the mean BER over five independent training and test runs and its error bar the minimum-to-maximum spread, where BER is the number of decoded bit errors over transmitted information bits. Each replicate uses $4\times10^6$ test bits by default, raised to $2.5\times10^8$ at $r_0=9.5~\mu\mathrm{m}$ and $5\times10^7$ at $M_0=600$.

Every threshold is optimized separately for each method, point, and replicate. Panels~(a)--(d) share a single SBRNN per method and loss, trained once on the pooled counts of all 32 channel points ($10^5$ message bits each) and reused across the four channel sweeps, whereas panels~(e)--(h) train a dedicated model at each point. Inputs are standardized by the training statistics, and we minimize the masked or unmasked cross-entropy with Adam (learning rate $10^{-3}$, batches of 500 windows). After each epoch, the loss is scored on a fixed held-out set of $5\times10^4$ message bits per point, pooled across points for the pooled models. Checkpoints require a relative improvement of at least 0.1\%, and training stops after 20 non-improving epochs or 4000 epochs, using the retained checkpoint. Both detectors are trained on the encodings of the same $10^5$ message bits. Since an RLIM$_2$ block spans $n>k$ channel uses, its stream admits about $n/k$ times as many sliding windows as the uncoded one. We uniformly subsample the coded windows to the uncoded count, so the two are matched in both message bits and number of training windows.

The reference MLSE is a reduced-state Viterbi detector~\cite{Forney1972MLSE,Eyuboglu1988RSSE} that keeps the first $L_{\mathrm{det}}$ channel taps, folds the remaining tail into a residual mean and variance, and scores branches with a Gaussian metric for the binomial arrivals and receiver noise; all panels use $L_{\mathrm{det}}=12$ except the capacity sweep of Fig.~\ref{fig:results}(g). There we sweep $L_{\mathrm{det}}\in\{6,8,10,12,14,17\}$ so the MLSE scales together with the SBRNN, and place both on a common storage axis: the SBRNN's trainable parameters against the MLSE's stored coefficients $3\cdot2^{L_{\mathrm{det}}}+L_{\mathrm{det}}$, from 198 to 393233. The thresholds, which store nothing, appear as horizontal bands, and the MLSE alone is given exact CSI.

First, masking never hurts: across the 57 matched SBRNN points it lowers the BER of the unmasked RLIM$_2$ detector at 56 and matches it at the remaining one, for a mean gain of $1.94\times$. Second, the coding gain survives neural detection: the best channel-blind RLIM$_2$ receiver beats the best channel-blind uncoded receiver at 40 of the 57 points by $5.29\times$ on average, and trails at the other 17 by only $1.89\times$. Allowing the CSI-aided MLSE on both sides narrows the margin but does not overturn it: using nearest-storage MLSE points in Fig.~2(g), coded transmission wins 31 of the 57 points (mean $3.77\times$) against 26 for uncoded (mean $1.88\times$).

Figs.~2(a)--(d) use SBRNNs trained jointly on all 32 channel points, while Figs.~2(e)--(h) train a dedicated SBRNN at each point. Placing the two molecule-budget sweeps side by side, Fig.~\ref{fig:results}(a) against Fig.~\ref{fig:results}(e), shows why this matters. At large $M_0$ the pooled coded-SBRNN curve flattens, because one cross-condition model cannot specialize to the easiest channels, whereas the per-point model keeps descending: at $M_0=600$ the masked RLIM$_2$-SBRNN improves from $8.94\times10^{-6}$ pooled to $6.48\times10^{-7}$ per point, an order of magnitude. The coding-plus-masking gain is therefore not an artifact of pooled training; specializing the detector to its operating point only widens it, so the pooled results understate what is achievable. Still, the pooled model is the practically important one: a single network, trained once and using no channel knowledge, holds the coding gain across all 32 conditions, the multi-scenario robustness for which the SBRNN was tested in~\cite{FarsadGoldsmith2018}.

 Fig.~\ref{fig:results}(f) places most of the context gain by $L\approx10$--$15$, so $L=40$ lies safely on the plateau. Fig.~\ref{fig:results}(g) compares storage: the masked RLIM$_2$-SBRNN stays between $3.84\times10^{-5}$ and $5.33\times10^{-5}$ across its whole range, from 242 to 363202 parameters ($3.84\times10^{-5}$ at the 2498-parameter default), whereas a storage-matched RLIM$_2$ MLSE needs 3082 coefficients merely to reach $6.51\times10^{-5}$ and still sits at $5.25\times10^{-5}$ with 393233, so the mask pays off most where the receiver must be smallest. Finally, Fig.~\ref{fig:results}(h) shows that the masked detector converges faster with training-set size than the unmasked one. 

Fig.~\ref{fig:results}(g) also compares the channel-blind masked SBRNN with the CSI-aided uncoded MLSE at matched storage ($M_0=400$). Using no channel knowledge, the masked SBRNN is the more accurate receiver across the compact and moderate range, from $3.7\times$ at 242 parameters to $1.15\times$ at 9090. The uncoded MLSE wins only at the largest size, by $1.26\times$ (393233 coefficients); beating the 2498-parameter default alone takes about $20\times$ the storage and exact CSI. A strong enough uncoded receiver can therefore overtake the coded one, which to our knowledge has not been reported for MC, but only outside the compact, CSI-free regime where such receivers are expected to operate.

\begin{table}[!t]
\centering
\caption{Diagnostics for the mask's assumption at $r_0=11.5~\mu\mathrm{m}$ and $r_0=9.5~\mu\mathrm{m}$ (five-run means; conditional BER excludes runs with no affected blocks). An affected block is one in which at least one triggering $1$-bit is missed. M: masked, U: unmasked.}
\label{tab:circularity}
\renewcommand{\arraystretch}{1.2}
\setlength{\tabcolsep}{5pt}
\footnotesize
\resizebox{\columnwidth}{!}{%
\begin{tabular}{lcccc}
\hline
& \multicolumn{2}{c}{hard ($r_0=11.5~\mu\mathrm{m}$)} & \multicolumn{2}{c}{easy ($r_0=9.5~\mu\mathrm{m}$)} \\
& M & U & M & U \\
\hline
Triggering-$1$ miss rate & $1.01\times10^{-2}$ & $1.87\times10^{-2}$ & $2.14\times10^{-8}$ & $9.02\times10^{-8}$ \\
False-$1$ rate at forced zeros & $4.40\times10^{-1}$ & $5.74\times10^{-3}$ & $2.18\times10^{-1}$ & $2.37\times10^{-9}$ \\
Fraction of affected blocks & $5.21\times10^{-2}$ & $9.24\times10^{-2}$ & $1.15\times10^{-7}$ & $4.86\times10^{-7}$ \\
BER in affected blocks & $2.90\times10^{-1}$ & $3.30\times10^{-1}$ & $2.73\times10^{-1}$ & $4.22\times10^{-1}$ \\
Overall BER & $2.42\times10^{-2}$ & $3.22\times10^{-2}$ & $1.34\times10^{-7}$ & $2.40\times10^{-7}$ \\
\hline
\end{tabular}%
}
\end{table}

The mask leaves each forced-zero window out of training because the decoder overwrites it, and the decoder does so only when the triggering $1$-bit before the window is detected. Table~\ref{tab:circularity} tests whether that holds, at a hard ($r_0=11.5~\mu\mathrm{m}$) and an easy ($r_0=9.5~\mu\mathrm{m}$) point spanning the BER range. The first row measures the assumption directly: the triggering-$1$ miss rate is small at the hard point and falls to $2\times10^{-8}$ at the easy one, so the bit the mask depends on is reliably detected, and almost never missed in the low-BER regime. The next rows show what each network learns at the forced zeros: the unmasked network drives their false-$1$ rate far below the masked network's, spending capacity to fit positions the decoder later overwrites. Even so, the masked network has the lower overall BER at both points: by missing fewer triggering $1$-bits it corrupts fewer blocks, and the error inside those blocks is no worse.

\section{Conclusion}
\label{sec:conclusion}
This letter has shown that a constrained code can reshape a neural receiver's learning objective: by removing the RLIM$_2$ positions the decoder overwrites, the mask has redirected compact-model capacity toward the triggering-$1$ and information-bearing positions. The RLIM coding gain previously established under threshold detection has survived neural detection, the masked RLIM$_2$-SBRNN beating the best uncoded receiver, threshold or SBRNN, at 40 of 57 points, and masking improving the unmasked detector at 56 of 57 with no loss. Using no channel knowledge, the compact masked detector has matched or beaten a storage-matched, CSI-aided reduced-state MLSE; only an uncoded detector with large storage and exact CSI has overtaken it, outside the compact, CSI-free regime where a nanoscale receiver is expected to operate. Future work will extend decoder-aware losses to broader constrained-code families, including reliability-weighted masking strategies. Future work will also investigate quantized implementations and experimental validation in molecular and Internet of Bio-Nano Things settings.

\bibliographystyle{IEEEtran}
\bibliography{references}

\end{document}